\documentclass[aps,prd,twocolumn,superscriptaddress,groupedaddress,nofootinbib]{revtex4-1}  

\usepackage[utf8]{inputenc}

\usepackage{graphicx}  
\usepackage{dcolumn}   
\usepackage{bm}        
\usepackage{amssymb}   
\usepackage{appendix}

\begin{document}

\widetext


\title{Localization with overlap fermions}

\author{ Réka Á. Vig}

\affiliation{University of Debrecen, H-4032 Debrecen, Bem tér 18/A, Hungary\\}

\author{Tamás G. Kovács} 

\affiliation{Eötvös Loránd University, H-1117 Budapest, Pázmány Péter sétány
  1/A, Hungary\\ and \\
 Institute for Nuclear Research, H-4026 Debrecen, Bem tér 18/c, Hungary}

\date{\today}
\begin{abstract}
We study the finite temperature localization transition in the spectrum of the
overlap Dirac operator. Simulating the quenched approximation of QCD, we
calculate the mobility edge, separating localized and delocalized modes in the
spectrum. We do this at several temperatures just above the deconfining
transition and by extrapolation we determine the temperature where the
mobility edge vanishes and localized modes completely disappear from the
spectrum. We find that this temperature, where even the lowest Dirac
eigenmodes become delocalized, coincides with the critical temperature of the
deconfining transition. This result, together with our previously obtained
similar findings for staggered fermions shows that quark localization at the
deconfining temperature is independent of the fermion discretization,
suggesting that deconfinement and localization of the lowest Dirac eigenmodes
are closely related phenomena.
\end{abstract}

\maketitle

\section{Introduction}
   \label{sec:introduction}

Strongly interacting matter is known to undergo a crossover at high
temperature. In the low temperature regime quarks are bound together to form
hadrons due to color confinement. During the crossover the boundaries of the
hadrons become blurred and matter goes into the state of quark-gluon plasma
(QGP). At the same time the spontaneously broken chiral symmetry becomes
approximately restored. Besides deconfinement and chiral restoration, there is
a third phenomenon that happens in the crossover region. Above the crossover
temperature the lowest lying eigenmodes of the Dirac operator become spatially
localized \cite{Halasz:1995vd}-\cite{Giordano:2014qna}. This is in sharp
contrast to the temperature regime below the crossover, where all the quark
eigenmodes are extended \cite{Verbaarschot:2000dy}.

In the high temperature phase the spectrum of the Dirac operator can be
separated into two regions. At the low end of the spectrum there are only
localized eigenmodes and their eigenvalues can be described by Poisson
statistics. In the upper part of the spectrum the eigenmodes are extended and
the corresponding eigenvalues obey Wigner-Dyson statistics
\cite{Verbaarschot:2000dy}. At fixed temperature this transition in the
spectrum between the localized and extended eigenmodes was shown to be a
genuine second order transition, and its correlation length critical exponent
was found to be compatible with that of the Anderson model in the same
symmetry class \cite{Giordano:2013taa}\footnote{Note, however, that in
  contrast to the Anderson transitions in condensed matter systems, in QCD
  this is not a genuine physical phase transition, as $\lambda$, the location
  in the Dirac spectrum is not a tuneable physical control
  parameter.}. Building on this analogy with Anderson transitions, we call the
critical point separating the localized and extended modes in the spectrum,
the mobility edge, $\lambda_c$ \cite{mobility}. While in the Anderson model
the mobility edge is controlled by the amount of disorder in the system, in
QCD an analogous role is played by the physical temperature. As the
temperature is lowered towards the crossover, the mobility edge moves down in
the spectrum, the part of the spectrum corresponding to localized eigenmodes
occupies a narrower and narrower band in the spectrum around zero.
Eventually, at a well defined temperature that we denote by $T_c^{loc}$, the
mobility edge vanishes, implying that even the lowest Dirac eigenmodes become
delocalized.

In QCD with physical quark masses the critical temperature of the localization
transition, $T_c^{loc}$ is in the crossover region \cite{Giordano:2014qna}.
This raises the question whether this is just a coincidence or there is some
deeper physical connection between the localization transition and the chiral
and/or the deconfinement transition. A possible way to test this is to move in
the parameter space of QCD to a regime where there is a genuine finite
temperature phase transition and check whether its critical temperature
coincides with $T_c^{loc}$. The simplest way to do that is to consider the
limit of infinitely heavy quarks, i.e.\ the quenched approximation to QCD,
which is known to have a first order deconfining phase transition at a
temperature of around $300$~MeV.

The possibility of linking the QCD transition to an Anderson-type localization
transition in the Dirac spectrum was first raised more than ten years ago by
Garcia-Garcia and Osborn. They studied the spectral statistics of the Dirac
operator in an instanton liquid model \cite{GarciaGarcia:2005vj} and in
quenched as well as full lattice QCD \cite{GarciaGarcia:2006gr} and found
evidence that around the chiral transition the spectral statistics of the
Dirac spectrum changes from Wigner-Dyson towards Poisson. This indicates that
the chiral transition is accompanied by a localization transition for the
lowest eigenmodes of the Dirac operator, however, at that time no no attempt
was made at a determination of $T_c^{loc}$, the critical temperature of the
localization transition, with a precision comparable to how $T_c$, the critical
temperature of the quenched deconfining phase transition is available in the
literature.

In a previous paper we explored this possibility by studying the spectrum of
the staggered quark Dirac operator in quenched gauge field backgrounds,
generated just above the finite temperature phase transition
\cite{Kovacs:2017uiz}.  For staggered fermions we calculated $T^{loc}_c$, the
critical temperature of the localization transition and found that it
coincided with that of the deconfining transition. Our results, obtained on
lattices with three different temporal extensions, $L_t=4, 6$ and 8, suggest
that the agreement of the localization and the deconfining transition
temperature is universal and is very likely to hold in the continuum limit,
provided the staggered discretization of quarks is in the correct universality
class also for the localization transition.

Unfortunately, staggered quarks are not in the same random matrix theory
symmetry class as continuum quarks \cite{Verbaarschot:2000dy}. Moreover, their
chiral symmetry is also different from that of continuum quarks. Although the
staggered Dirac operator is expected to have the correct continuum limit, it
is still possible that at finite lattice spacing it does not properly
describe some properties of the lowest quark eigenmodes, the ones that are our
main concern here for studying the localization transition. This is a
potentially important issue, as the lowest part of the Dirac spectrum is
particularly sensitive to the chiral properties of the given
discretization. Therefore, in the present work we chose to repeat our previous
staggered study with the overlap Dirac operator that has exact chiral symmetry
already for finite values of the lattice spacing
\cite{Narayanan:1993ss}.

Besides this, there are two more reasons concerning localization, why it is
important to verify our previous results with the overlap. Firstly, overlap
fermions with the $SU(3)$ gauge group are in the same random matrix symmetry
class, the chiral unitary class, as fermions in the continuum. Secondly,
unlike the staggered action that is ultralocal, the overlap action couples
quark degrees of freedom to arbitrarily large distances, albeit with couplings
falling exponentially with the distance. Since in the theory of Anderson-type
models localization is generally known to strongly depend on the range of the
couplings (hopping terms in the Hamiltonian) \cite{Evers:2008zz}, it is
interesting to check whether the non-locality of the overlap Dirac operator
has any influence on the localization transition in QCD. In fact, to our
knowledge, this is the first study where the mobility edge is explicitly
determined in QCD with chiral quarks\footnote{Indirect evidence for
  localization of overlap quarks has already been obtained by studying the
  distribution of the lowest two eigenvalues in Ref.\ \cite{Kovacs:2009zj},
  but the transition to the delocalized regime in the spectrum was not
  explicitly seen in that work.}.

In the present work we used a subset of the gauge configurations that were
previously generated for our earlier staggered study. Since overlap spectra
are significantly more expensive to calculate than staggered spectra, here we
limited our study to one value of the temporal lattice size, $L_t=6$. We
computed the mobility edge for gauge ensembles generated with six different
values of the gauge coupling, $\beta$, all corresponding to temperatures
slightly above the deconfining transition. By extrapolation we determined the
gauge coupling $\beta_c^{loc}$ where the mobility edge vanished and all
localized eigenmodes disappeared from the Dirac spectrum. Confirming our
previous staggered result, we found $\beta_c^{loc}$ to be compatible with the
critical gauge coupling of the deconfining phase transition for $L_t=6$.

The plan of the paper is as follows. In Sec.\ \ref{sec:mobility_edge} we
describe the lattice ensembles used for the calculation and show how we
computed the mobility edge from the Dirac spectra. In
Sec.\ \ref{sec:loc_transition} we discuss the determination of the critical
coupling of the localization transition. In Sec. \ref{sec:conclusions} we draw
our conclusions and finally in the Appendix we describe the technical details
of the unfolding of the spectrum.

\section{Calculation of the mobility edge}
  \label{sec:mobility_edge}

The Dirac operator that we used for this study was the overlap with Wilson
kernel parameter $M=-1.3$. As smearing of the gauge field is known to improve
some properties of the overlap and also makes the calculations faster
\cite{Kovacs:2002nz}, two steps of hex smearing \cite{Capitani:2006ni} were
applied to the gauge field before inserting it into the overlap. The gauge
field configurations we used here were quenched Wilson action lattices with
temporal extension $L_t=6$.  In Table~\ref{tab:parameters} we collected the
parameters of the simulations.

\begin{table}
 \begin{ruledtabular}
 \begin{tabular}{lrrr}
   $\beta$ & $L_s$ & $N_c$ & $N_{evs}$ \\ \hline
     5.91   & 40    &  741  & 80 \\
      5.92  & 40    &  821  & 80 \\
		&32 & 3823  & 50 \\
		&24 & 4668  & 25 \\
      5.93  & 40    &  750  & 80 \\
      5.94   & 40   &  856  & 80 \\
      5.95  & 40    &  835  & 80 \\
      5.96 & 40     &  609  & 80 \\
           & 24     & 3915  & 25 \\
 \end{tabular}
 \end{ruledtabular}
\caption{\label{tab:parameters} Simulation parameters (from left to right):
  the Wilson plaquette gauge coupling, the size of the lattice in the spatial
  direction, the number of configurations and the number of eigenvalues
  computed for each configuration. All the lattices had a temporal extension
  of $L_t=6$.}
\end{table}

On each gauge configuration we computed a number of lowest eigenvalues of
$D^\dagger D$, where $D$ is the overlap Dirac operator. In what follows we
always work with the eigenvalues of $D^\dagger D$ that are the magnitude
squared of the corresponding eigenvalues of the Dirac operator $D$. Since our
analysis is based on the unfolded spectrum, which is invariant with respect to
monotonic reparametrizations of the spectrum, it makes no difference that we
perform the analysis in terms of the eigenvalues of $D^\dagger D$. As
explained in the Appendix, we take extra care to make even the assignment of
eigenvalue pairs to spectral windows to be reparametrization invariant. To
make the notation simpler and avoid having to write the absolute value squared
everywhere, we denote by $\lambda$ the eigenvalues of $D^\dagger D$, in terms
of which we perform the entire analysis.

The number of eigenvalues to be computed per configuration was chosen to
include all the eigenvalues to a point well above the mobility edge,
$\lambda_c$\footnote{This criterion could be checked only a posteriori, after
  determining $\lambda_c$}. Having exact chiral symmetry, the overlap
possesses exact zero eigenvalues in gauge field backgrounds with non-zero
topological charge. Since these eigenvalues are all exactly at the lower edge
of the spectrum, they do not contain any information relevant to the present
study, we simply removed them from the spectra before further analysis.

Localized and delocalized eigenmodes are characterized by different statistics
of the corresponding eigenvalues. To track the transition throughout the
spectrum and locate the mobility edge, we used the simplest spectral
statistics, the unfolded level spacing distribution (ULSD), calculated
locally, within narrow spectral windows of the spectrum. Unfolding, a
transformation well known in the theory of random matrices, is a monotonic
mapping of the spectrum that sets the local spectral density to unity
everywhere throughout the spectrum. In particular, by construction, the
unfolded eigenvalues are dimensionless and their average level spacing is
unity. More details on how the unfolding was done are presented in the
Appendix.

Unfolding is useful since both for localized and delocalized eigenmodes,
universally valid analytic results are known for the ULSD of the corresponding
eigenvalues \cite{Verbaarschot:2000dy}. Spectra corresponding to localized
eigenmodes obey Poisson statistics and the ULSD follow the exponential
distribution,
\begin{equation}
  p(s)=\exp(-s),
      \label{eq:exp}
\end{equation}
where $s$ is the level spacing between the nearest neighbor unfolded
eigenvalues.  

\begin{figure}
\centering
\includegraphics[width=1\columnwidth]{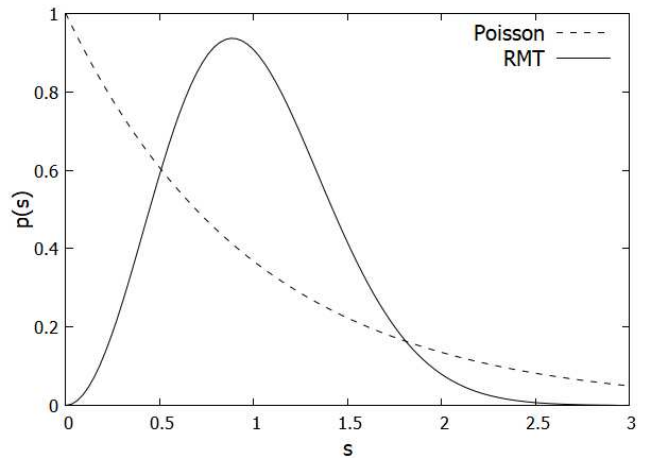}
\caption{\label{fig:poisson_wigner}The probability density functions of the
  level spacings in the cases when the eigenvalues obey the Poisson statistics
  (exponential distribution, dashed line) and Wigner-Dyson statistics
  (continuous line, Wigner surmise).}
\end{figure}

In the case of extended modes the ULSD is also known analytically, however, it
is much more complicated than in the localized case and also depends on the
random matrix symmetry class of the given model. A very good approximation to
the ULSD in this case is provided by the so called Wigner surmise that for the
unitary symmetry class, to which the overlap operator belongs, reads as
\begin{equation}
  p(s)= \frac{32}{\pi ^2}s^2 \exp(-  \frac{4}{\pi}s^2).
     \label{eq:ws}
\end{equation}
Notice that both the exponential and the Wigner surmise distribution are
universal in the sense that they are free of any adjustable parameters. In
particular, the originally dimensionful parameter, the local spectral density
has been removed from the spectrum by the unfolding. For further reference we
plotted the two distributions in Fig.~\ref{fig:poisson_wigner}.

\begin{figure}
\centering
\includegraphics[width=0.7\columnwidth]{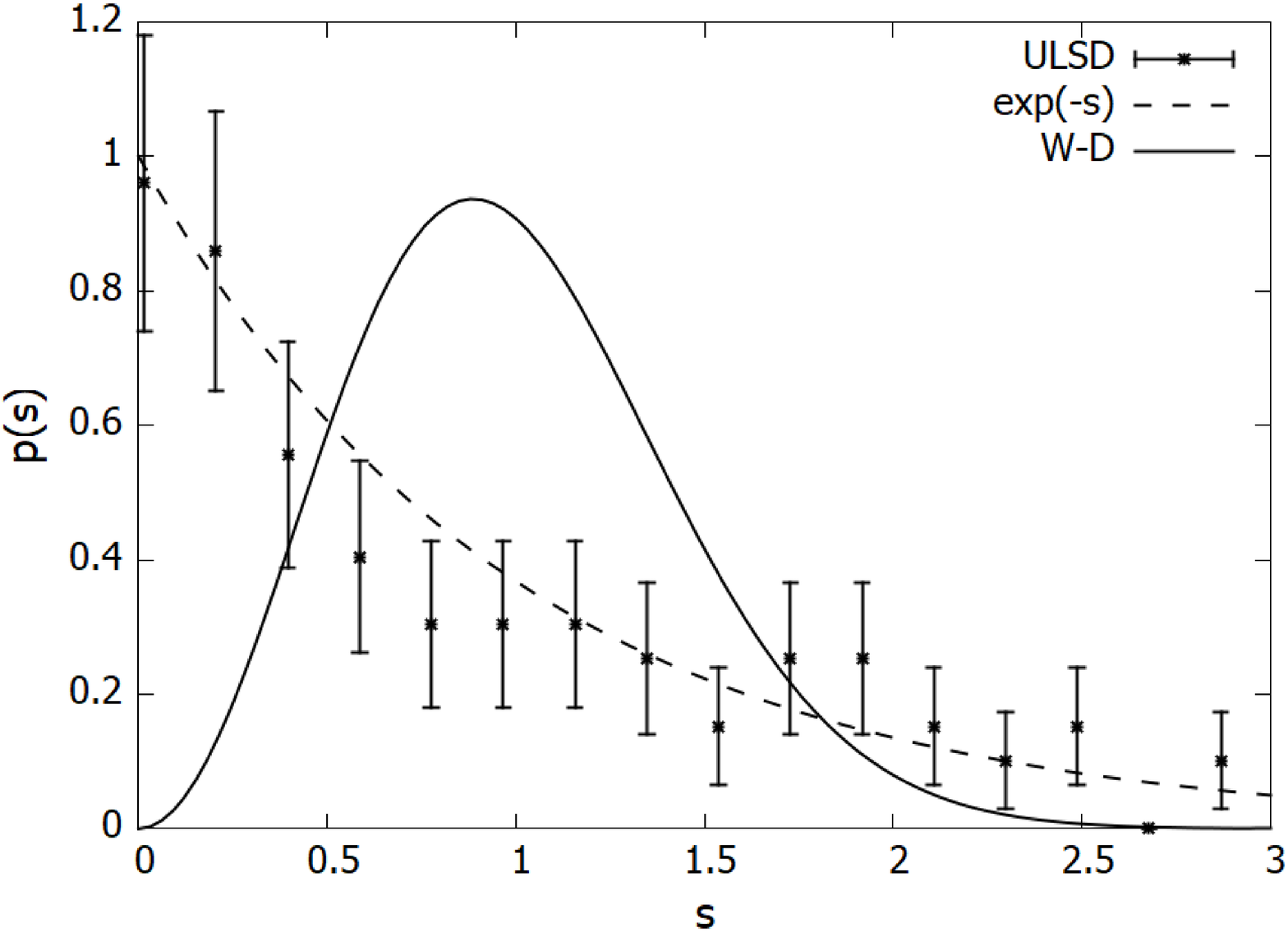}
\includegraphics[width=0.7\columnwidth]{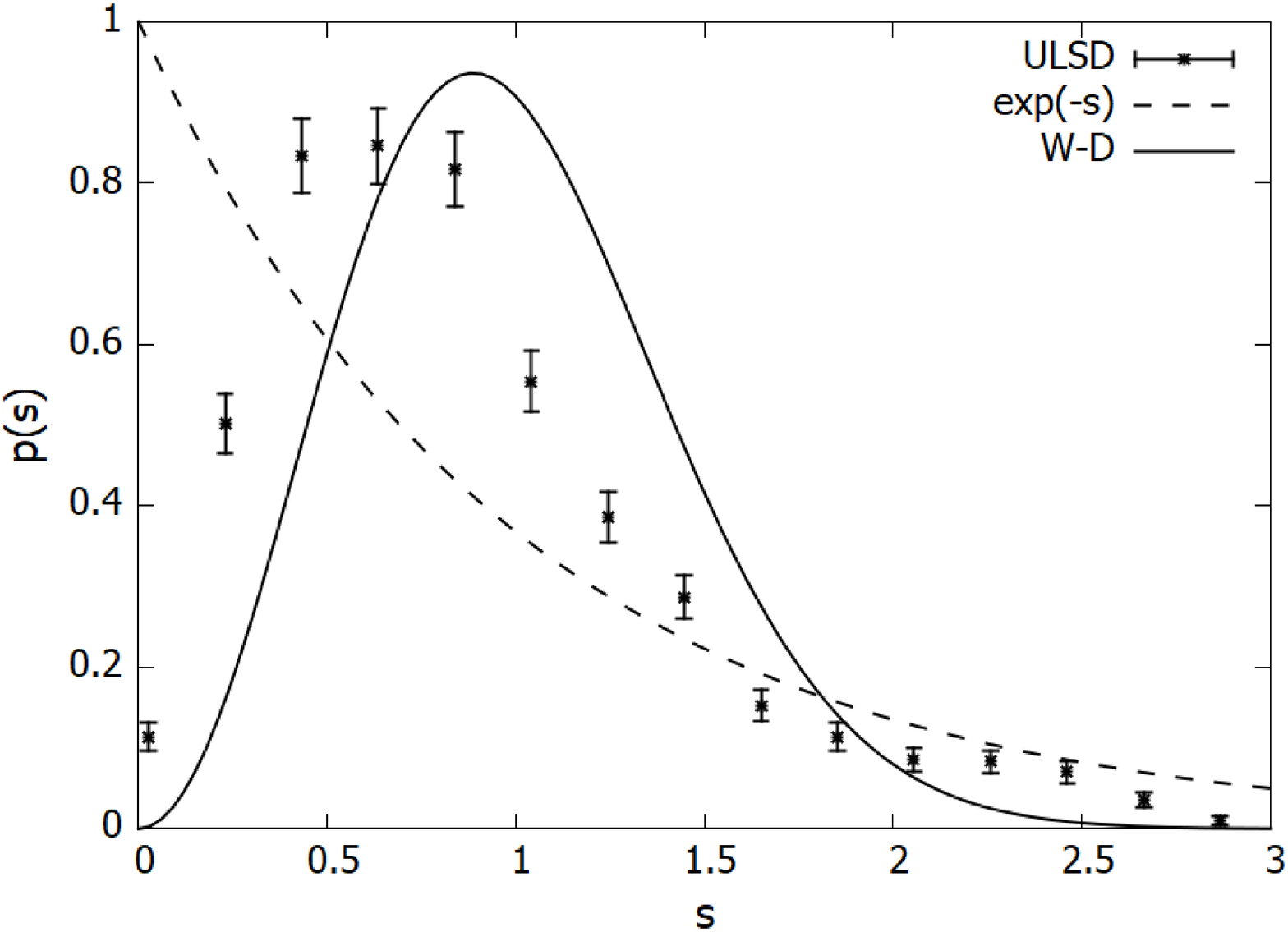}
\includegraphics[width=0.7\columnwidth]{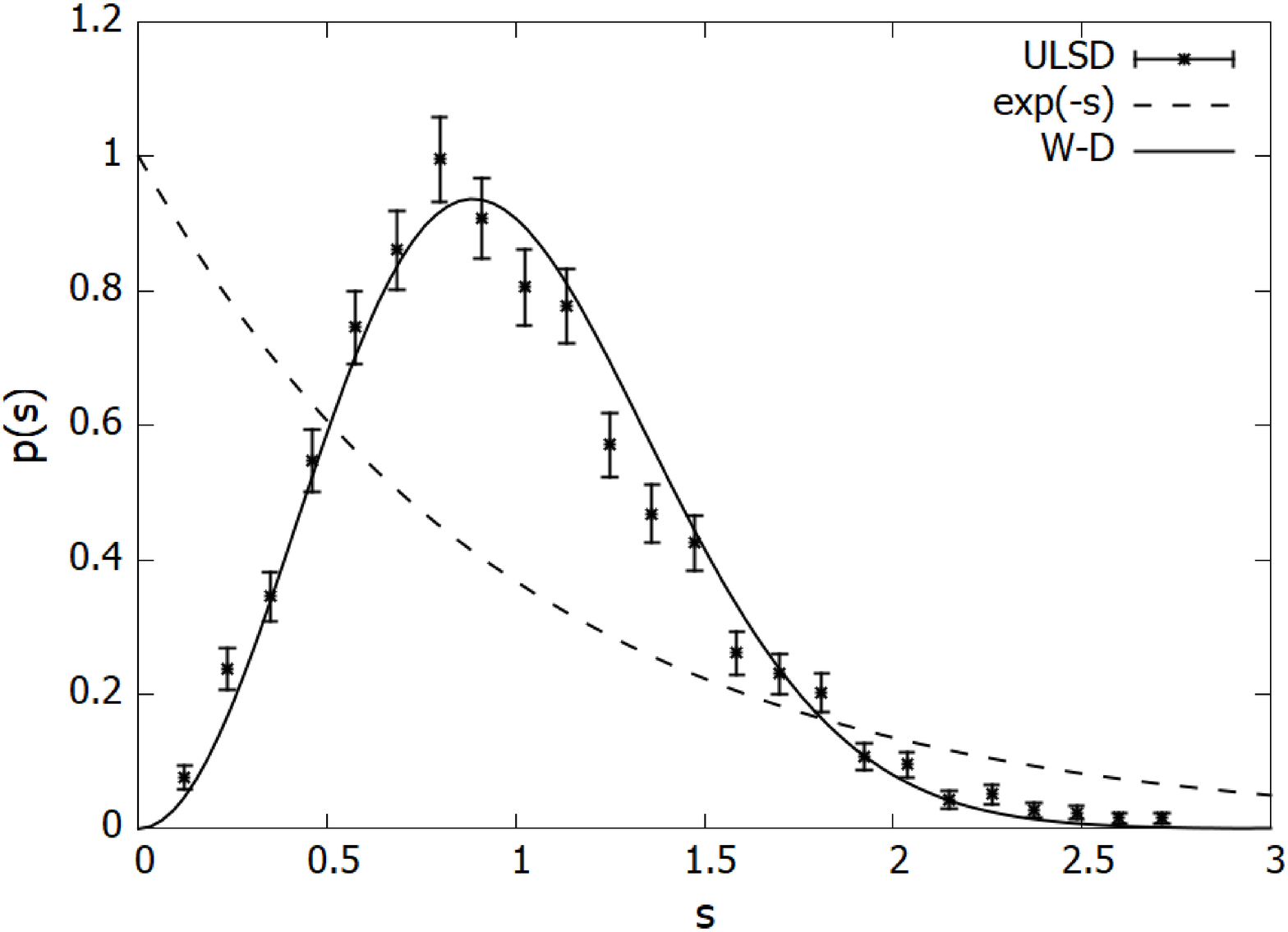}
\caption{\label{fig:transition}The unfolded level spacing distribution in
  three different spectral windows for the $\beta=5.95$, $L_s=40$
  ensemble. The spectrum is scanned from the lowest eigenvalues (top panel)
  with Poisson statistics, through the critical region (middle panel), up to
  the regime with Wigner-Dyson statistics (bottom panel). We also show the
  expected limiting distributions, the exponential (dashed line) and the
  Wigner surmise (continuous line).}
\end{figure}

Our aim here is to scan the spectrum starting from zero and follow how the
{\it local} ULSD changes from the exponential distribution of
Eq.\ (\ref{eq:exp}) to the Wigner surmise of Eq.\ (\ref{eq:ws}). To this end
we divide the spectrum into narrow spectral windows, compute the ULSD
separately in each spectral window and follow how it changes throughout the
spectrum. In Fig.\ (\ref{fig:transition}) we show how the unfolded level
spacing distribution evolve as the spectrum is scanned starting from the
lowest eigenvalues (top panel) crossing the critical, transition region
(middle panel) and finally moving up to the Wigner-Dyson regime (bottom panel).

However, monitoring the continuous change of a function (here the
probability density of the unfolded level spacings) is complicated. To make
this task easier, we choose a single parameter of this distribution and
monitor how that changes throughout the spectrum. A simple choice for this
parameter is the integral
\begin{equation}
  I_{s_0}=\int_0^{s_0} p(s) ds 
    \label{eq:is0}
\end{equation}
of the probability density up to the lowest crossing point $s_0 \approx
0.508$ of the two limiting distributions, the exponential and the Wigner
surmise. This choice of $s_0$ has the advantage that it maximizes the
difference of the integral between the two limiting cases and thereby
facilitates their clear separation. 

\begin{figure}
\centering
\includegraphics[width=1\columnwidth]{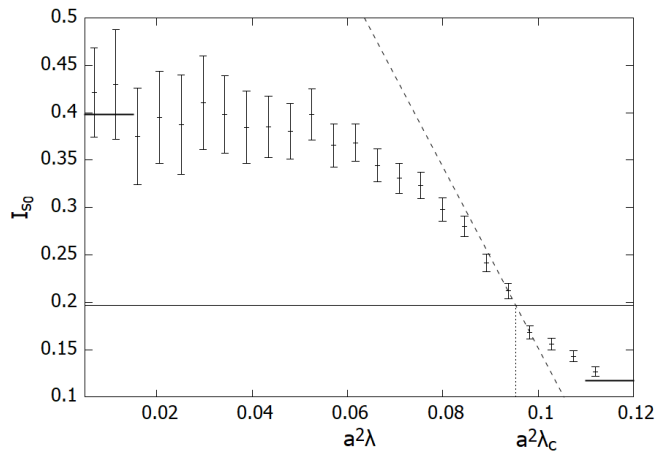}
\caption{\label{fig:is0}The integrated probability density (defined in
  Eq.~\ref{eq:is0}) as a function of eigenvalues $\lambda$ of $D^\dagger
  D$. The figure shows the data for $\beta = 5.95 $ with a spatial volume of
  $40^3$. The two short horizontal lines in the top left and the bottom right
  corner of the figure indicate the limiting values of $I_{s_0}$ for the
  Poisson (localized) and the Wigner-Dyson (delocalized) statistics. }
\end{figure}

An example of how $I_{s_0}$ changes through the spectrum is shown in
Fig.~\ref{fig:is0}. As expected and can also be seen in the figure, in a
finite volume $I_{s_0}$ changes smoothly from the value $I_{s_0}^P\approx
0.398$ corresponding to the exponential distribution to $I_{s_0}^{W} \approx
0.117$ corresponding to the Wigner surmise. However, based on the finite size
scaling study of Ref.~\cite{Giordano:2013taa}, in the thermodynamic limit we
expect the transition to become singular, as in a second order phase
transition. The mobility edge, $\lambda_c$ that we eventually want to locate
is this sharply defined singular transition point appearing only in the
infinite volume limit. In a finite volume the definition of the ``critical
point'' is somewhat arbitrary, however, a good choice is the point in the
spectrum where $I_{s_0}$ is equal to the value $ I_{s_0}^{crit}=0.1966$,
corresponding to the critical distribution, known from the finite size scaling
study of Ref.~\cite{Giordano:2013taa}. From now on, with a slight abuse of
notation, we will call the point in the spectrum, $\lambda_c$, for which
$I_{s_0}(\lambda_c)=I_{s_0}^{crit}$, the mobility edge.

The quantity $\lambda_c$, defined in this way, can still have a volume
dependence, but it is a good approximation to the mobility edge in the
thermodynamic limit. To keep the finite size corrections under control, we
calculated $\lambda_c$ on lattices of spatial linear size $L_s=24, 32,
40$. While the results on the smallest volume differed significantly from
those on the other volumes, the results from the larger two volumes agreed
within the statistical uncertainties. Therefore, for the rest of the analysis
we always used the data from the largest volume, $L_s=40$.

Since the function $I_{s_0}(\lambda)$ has an inflection point at $\lambda_c$,
around this point it can be well approximated with a straight line.  We could
thus easily determine $\lambda_c$ by solving the equation
$I_{s_0}(\lambda_c)=I_{s_0}^{crit}$ by approximating the function
$I_{s_0}(\lambda)$ with a linear fit to the data in the given range (see
Fig.~\ref{fig:is0}).

\section{The critical temperature of the localization transition}
    \label{sec:loc_transition}

\begin{figure}
\centering
\includegraphics[width=1\columnwidth]{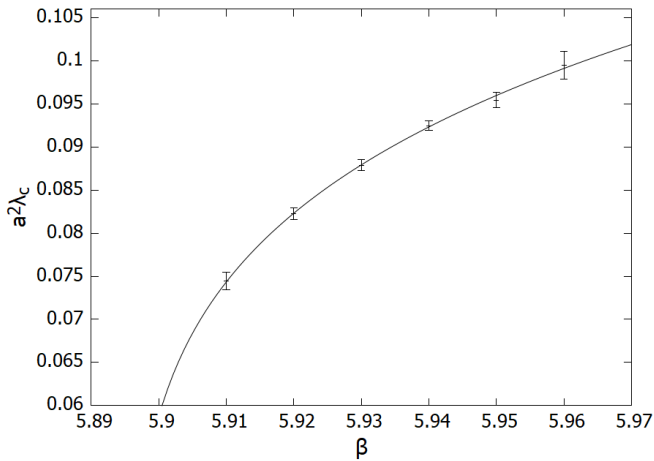}
\caption{\label{fig:lc_vs_beta} The mobility edge as the function of the
  gauge coupling, approximated with a power function.}
\end{figure}

So far we have shown how to calculate the mobility edge, $\lambda_c$, at a
given temperature. Our final goal is to determine the temperature where the
mobility edge vanishes and localized modes completely disappear from the Dirac
spectrum. Since we keep the temporal size of the lattice fixed, the
temperature can be controlled by the gauge coupling, $\beta$. We computed
$\lambda_c$ for lattice ensembles generated at several different values of the
gauge coupling, above, but close to the deconfining phase transition. The
results are shown in Fig.~\ref{fig:lc_vs_beta}. The range of couplings we used
were limited by two factors. Firstly, even though the deconfining transition
is of first order, the correlation length increases substantially towards the
transition which puts a lower limit to the couplings for which finite size
corrections can be kept under control. Secondly, we would like to extrapolate
the function $\lambda_c(\beta)$ to find where it vanishes, and for the
extrapolation only points close enough to the zero of this function are
useful. Since we expected the zero of the function $\lambda_c(\beta)$ to be
close to the deconfining transition, $\beta_c$, we limited our simulations to
couplings not too far from this point. 

Finally, for the extrapolation we used the ansatz
\begin{equation}
  \lambda_c (\beta) = p_1(\beta-\beta_c^{loc})^{p_2}, 
\end{equation}
and its parameters $p_1, p_2$ and $\beta_c^{loc}$ were fitted to the data. The
ansatz turned out to describe the data remarkably well and using all six data
points the resulting $\chi^2$ per degree of freedom was $\chi^2= 0.67$. The fit along with the data is shown in Fig.~\ref{fig:lc_vs_beta}.
The resulting location of the localization transition is
$\beta_c^{loc}=5.893(7)$, where we quoted the statistical uncertainty. Within
the uncertainties this agrees with the critical point of the deconfining
transition, $\beta_c=5.8943(3)$ \cite{Francis:2015lha}.  This, together with
similar results obtained in Ref.~\cite{Kovacs:2017uiz} with staggered
fermions, strongly suggest that independently of the fermion discretization,
the localization transition and deconfinement happen at the same temperature,
therefore the two phenomena are very likely to be strongly related.

\section{Conclusions}
   \label{sec:conclusions}

We examined the localization transition of the quarks using the quenched
approximation. We computed the lowest lying eigenvalues of the overlap Dirac
operator above the critical coupling of the deconfining transition. By
calculating the mobility edge, $\lambda_c$, for different gauge couplings we
determined the function $ \lambda_c (\beta) $ and extrapolated it to locate
$\beta_c^{loc}$, where the mobility edge vanishes and all the eigenmodes
become delocalized. We compared our result with the critical coupling of the
deconfining phase transition and found that the two critical couplings are
compatible; the localization transition and deconfinement occur at the same
temperature. This is in agreement with our previous similar results with
staggered fermions and indicates that localization and deconfinement are
strongly related phenomena.

The present work was motivated by the fact that in QCD with physical dynamical
quarks the localization transition occurs in the crossover region. On the one
hand, our results clearly indicate that the localization transition is
strongly related to deconfinement, which -- at least on a qualitative level --
probably carries over from the quenched model to real physical QCD. On the
other hand, the quenched model cannot properly account for the other important
transition, the chiral transition that also occurs in the QCD crossover. To
see how localization is related to chiral restoration, it would be interesting
to consider the other limiting case, the chiral limit. For massless light
quarks, the chiral transition is expected to become a genuine phase transition
\cite{Pisarski:1983ms} and it could be tested whether its critical temperature
agrees with the critical temperature of the localization transition. Although
simulations in the chiral limit are technically immensely challenging, such a
study could also provide additional insight into the physics of the
restoration of chiral symmetry, how that happens in the massless (chiral)
limit. Several questions related to this are currently under active study
\cite{Ding:2019prx}.

\begin{acknowledgments}
  
T.G.K.\ was partially supported by the Hungarian National Research,
Development and Innovation Office - NKFIH Grant No.\ KKP126769. R.A.V.\ was
partially supported by the New National Excellence Program of the Hungarian
Ministry for Innovation and Technology Grant No.\ ÚNKP-19-3-I-DE-490.
T.G.K. thanks Matteo Giordano, Sándor Katz and Dániel Nógrádi for helpful
discussions.

\end{acknowledgments}

\appendix*
\section{Unfolding} \label{unfolding}
   \label{sec:appendix}

Unfolding is a monotonic mapping of the eigenvalues that - by definition -
renders the spectral density unity throughout the unfolded spectrum. This
transformation is useful since it removes the scale, specific to the given
spectrum and reveals universal spectral fluctuations. In principle unfolding
can be done in several different ways, all equivalent for a dense enough
spectrum. Here we did the unfolding by taking all the eigenvalues from all the
configurations of the given ensemble and putting them into ascending order
according to their magnitudes. To each eigenvalue we assigned its rank divided
by the number of configurations, $N_c$, we used this mapping to define the
unfolded spectrum. In this way the level spacing between successive unfolded
eigenvalues is exactly $1/N_c$, which means that there are $N_c$ eigenvalues
in an interval of unit length anywhere in the unfolded spectrum. This implies
that the average spectral density per configuration is unity throughout the
unfolded spectrum.

In the present work we used the unfolded level spacing distribution (ULSD)
calculated from the spectrum unfolded in the above described way. In
particular, we followed how the ULSD changed throughout the spectrum, starting
from the Poisson statistics and going over to Wigner-Dyson statistics. This
required the calculation of the local ULSD at different locations in the
spectrum. In order to do this, we divided the spectrum into small spectral
windows and calculated the ULSD in each window separately.

In principle, this method is straightforward, if the spectrum is infinitely
dense. However, for finite density, there is an ambiguity in how we decide
whether a pair of neighboring eigenvalues belongs to the given spectral window
or not. We could demand that both members of the pair be within the spectral
window in question. However, this would artificially limit the largest
possible level spacings, especially for eigenvalues close to the edge of a
spectral window. To avoid this uncontrolled truncation of the tail of the ULSD
we chose the criterion that a pair of nearest neighbor eigenvalues was
considered to belong to the given spectral window if the midpoint of the pair
was in the window. To ensure that our procedure, including the assignment of
pairs to spectral windows, is invariant with respect to monotonic
reparametrizations of the spectrum, we applied the midpoint rule in the unfolded
spectrum. This is easily done by mapping the endpoints of the spectral window
into the unfolded spectrum. Notice, however, that we can and do still plot the
results in terms of the original (not the unfolded) spectrum, as seen in
Fig.\ \ref{fig:is0}.

\end{document}